\documentclass{article}
\usepackage{spconf,amsmath,graphicx}
\usepackage{multirow}
\usepackage{threeparttable}
\usepackage {setspace}
\usepackage{caption}
\title{More than Vanilla Fusion: a Simple, Decoupling-free, Attention Module for Multimodal Fusion Based on signal Theory}
\name{Peiwen Sun, Yifan Zhang, Zishan Liu, Donghao Chen, Honggang Zhang}
\address{Beijing University of Posts and Telecommunications}
\begin{document}
%\ninept
%
\maketitle
\begin{abstract}
The vanilla fusion methods still dominate a large percentage of mainstream audio-visual tasks. However, the effectiveness of vanilla fusion from a theoretical perspective is still worth discussing. Thus, this paper reconsiders the signal fused in the multimodal case from a bionics perspective and proposes a simple, plug-and-play, attention module for vanilla fusion based on fundamental signal theory and uncertainty theory. In addition, previous work on multimodal dynamic gradient modulation still relies on decoupling the modalities. So, a decoupling-free gradient modulation scheme has been designed in conjunction with the aforementioned attention module, which has various advantages over the decoupled one. Experiment results show that just a few lines of code can achieve up to 2.0\% performance improvements to several multimodal classification methods. Finally, quantitative evaluation of other fusion tasks reveals the potential for additional application scenarios.
\end{abstract}
\vspace{-0.4cm}
\section{Introduction}
\label{sec:intro}

\vspace{-0.2cm}

In recent years, we have witnessed a large number of multimodal methods \cite{Mo_2023_CVPR,multi4,qian2021audio,chen2020multi}, which comprehensively surpassed unimodal methods in multiple tasks. In in-depth investigations of the audio-visual area, some studies addressed imbalanced optimization and coverage rate as an obstacle in coarse-grained multimodal learning \cite{multi1,multi2}. Thus, the potential of multiple modalities could not be fully exploited \cite{multi3}. To cope with this, extra uni-modal classifiers \cite{multi2} and auxiliary loss \cite{multi2,multi5} were introduced to cultivate better models. Furthermore, studies \cite{multi2,xiao2020audiovisual} have shown that gradient modulation could alleviate the imbalance between modalities. As an effective modulation method, on-the-fly gradient modulation \cite{multi3} based on approximate uni-modal performance was proposed to enable parameters of different modalities to update with inconsistent learning rates. Then, the attention-based multi-modal fusion strategies \cite{multi4}, allowing cross-modal information interaction during training, alleviated the imbalance via information sharing.

The application of informatics knowledge has been utilized in the field of deep learning for a long time. Informatics correlation was first applied to the training loss of learning models based on canonical correlation analysis (CCA) \cite{sun2020learning,chang2018scalable}. As a competitive substitution, researchers \cite{wang2019efficient,lee2022maximal,liang2021person} proposed the deep learning implementation of HGR maximal correlation, Soft HGR, which has been proven to learn better correlation and to hold lower computing complexity than CCA. However, since the mechanism of auditory \cite{shinn2008object} and visual attention \cite{carrasco2011visual} was well demonstrated in cognitive science, another group of inference methods \cite{guo2022attention} based on bionic signal theory has emerged and has been proven to be enlightening in the field. SimAM \cite{energy1} proposed to treat the weight of neurons as signals and found the importance of each neuron by optimizing an energy function without adding extra parameters, which was found also suitable for speech-related tasks \cite{energy2}. Naturally, such an approach based on informatics energy theory can be utilized for audio and visual separately, so \textit{why not fusion}?
% Besides designing sophisticated blocks, another line of research focuses on building plug-and-play modules that can refine convolutional outputs within a block and enable the whole network
% to learn more informative features.

The “vanilla” fusion method, which simply consists of the regular transformation applied to multimodal inputs, can be described as a series of discrete signals undergoing simple operations. So, inspired by bionic thoughts, closed-form solutions based on signal theory are derived on the basis of SimAM \cite{energy1} to optimize energy and mutual energy. Based on the study \cite{uncertainty}, we find that this optimization can be understood from both an energy perspective and an uncertainty perspective, which further proves the feasibility of the theory. Furthermore, based on derivation, the vanilla fusion methods have been improved through just a few dozen lines of code, which is named \textit{Simple Attention Module for Multimodal} (SimAM$^2$). Then to avoid cumulative errors caused by uncoupling, we apply a method that does not require the decoupling of fusion features to achieve convergence supervision of different modalities. During the experiment, the methods above are individually validated for effectiveness and work together to yield up to 2.0\% improvement in results across multiple audio-visual classification datasets including CREMA-D \cite{dataset1}, VGGSound \cite{dataset2}. Finally, extra experiments beyond classification are elicited to demonstrate the possibility of other scenarios.

\begin{figure*}[t]
\centering
\includegraphics[width=14cm]{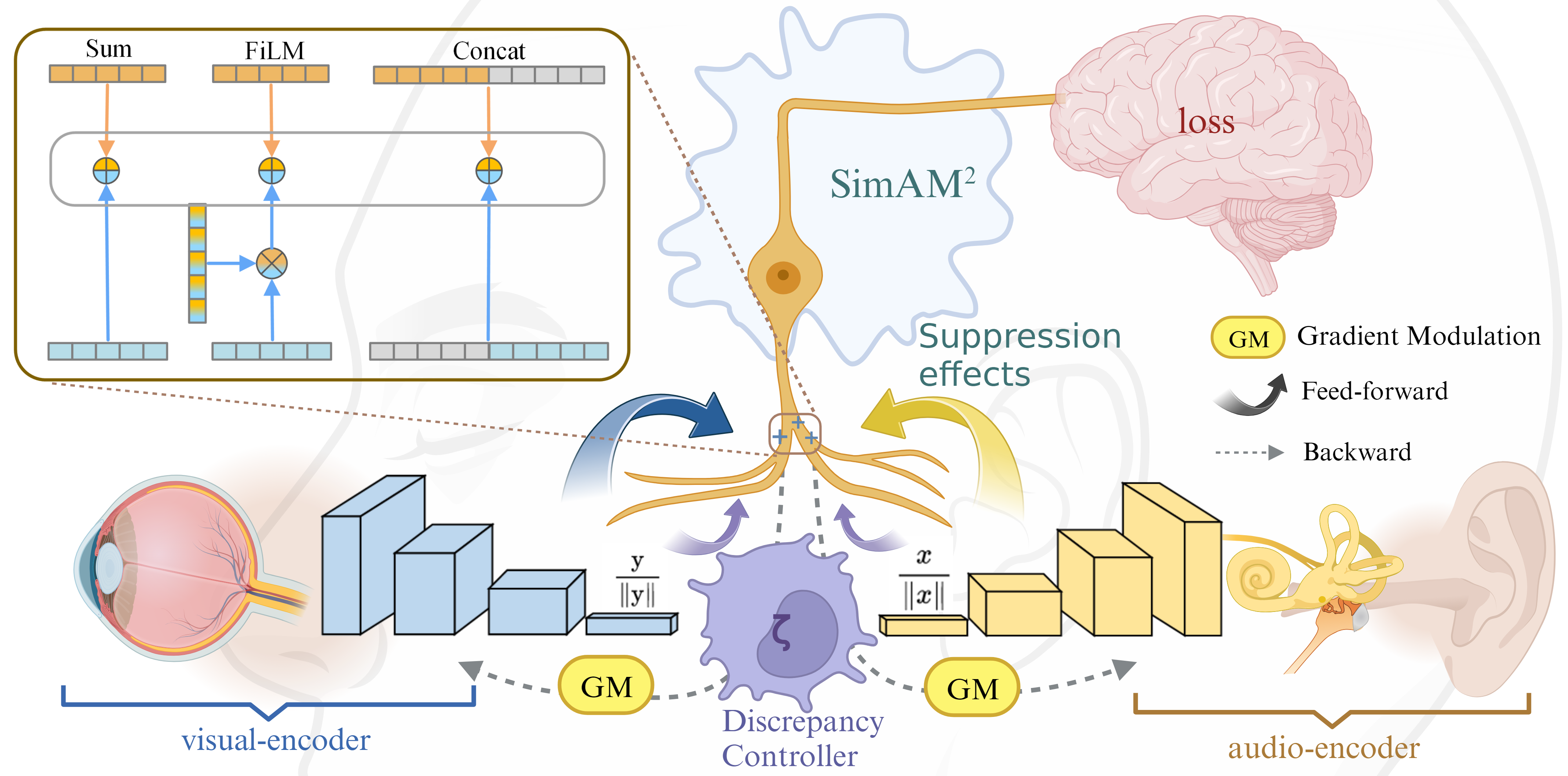}
\vspace{-0.3cm}
\captionsetup{font={small}} 
\caption{
\small
The pipeline of SimAM$^2$ in vanilla fusion} \label{fig:fusion_net}
% 左上引到右下
% decouple-free
% 
\vspace{-0.7cm}  %调整图片与上文的垂直距离
\end{figure*}

\vspace{-0.5cm}

\section{Method}
\label{sec:method}

\vspace{-0.25cm}

Here, two important assumptions are introduced. Firstly, in visual neuroscience, the most informative neurons are usually the ones that show distinctive stimulation patterns from surrounding neurons \cite{energy1}. In other words, the neurons displaying clear spatial \textbf{suppression effects} should be given higher priority. Moreover, the same thing goes on in multimodality processing. 
% When stimuli of different modalities act on the same neuron, the neuron is supposed to exhibit suppression effects as if it were unimodal. 
Secondly, when the current point is below the threshold of triggering the action potential, all electrical signals generated by nerve cells are \textbf{collectively superimposed} \cite{PNS} as Eq.\ref{con:summation} on the resting potential.
So, with multimodal stimuli, neurons need to make not only unimodal weight adjustments based on unimodal signal but also comprehensive adjustments based on the overall stimuli.

Therefore, we propose a series of modifications to vanilla fusion based on the assumption of linear superposition of neuronal stimuli and make adaptations to the overall training method. The elaboration on the overall method is divided into 3 parts, \textit{Simple Attention Module for Multimodal} in Sec.\ref{sec:att}, modification on vanilla fusion in Sec.\ref{sec:fusion}, and coupling-free approach in Sec.\ref{sec:Non-decoupled} to control training equilibrium.

\vspace{-0.3cm}

\subsection{Simple Attention Module for Multimodal (SimAM$^2$)}
\label{sec:att}
\vspace{-0.1cm}
% 多模态两个信号相加的推导
% 对sum和gated的改进
Since the spatial suppression effects assumption in unimodal neurons is exactly the same as SimAM \cite{energy1}, we follow most of the derivations of SimAM regarding the energy part of neurons. So the energy function of neurons still follows it as
\vspace{-0.3cm}
\begin{equation}
\begin{aligned} e_t\left(w_t, b_t, \mathbf{y}, x_i\right) = & \frac{1}{M-1} \sum_{i=1}^{M-1}\left(-1-\left(w_t x_i+b_t\right)\right)^2 \\
& +\left(1-\left(w_t t+b_t\right)\right)^2 +\lambda w_t^2.\end{aligned}
\vspace{-0.2cm}
\end{equation}
Note, $t$ and $x_i$ are the target neuron and other neurons; $\left(w_t x_i+b_t\right)$ and $\left(w_t t+b_t\right)$ are linear transforms of $x_i$ and $t$; $M$ is the number of neurons on that channel $\lambda$ is the regularization term coefficient. Here, $\lambda$, as is used to limit the growth of the model, should be discussed later. The same fast closed-form solution with respect to $w_t$ and $b_t$ is

\vspace{-0.2cm}
\begin{equation}
\vspace{-0.2cm}
\begin{aligned}
w_t &=\frac{2\left(t-\mu_t\right)}{\left(t-\mu_t\right)^2+2 \sigma_t^2+2 \lambda},\\
b_t &=-\frac{1}{2}\left(t+\mu_t\right) w_t.
\end{aligned}
\end{equation}
while $\mu_t$ and $\sigma_{t}^2$ are mean and variance calculated over all neurons $x_i$ except $t$ in that channel. The minimal energy can be expressed as a function with respect to $\lambda$ 
% Note $\lambda$ is set as a hyperparameter for further use.
\vspace{-0.2cm}
\begin{equation}
\vspace{-0.2cm}
e_t^{*}(\lambda)=\frac{4\left(\hat{\sigma}^2+\lambda\right)}{(t-\hat{\mu})^2+2 \hat{\sigma}^2+2 \lambda},
\label{con:regu}
\end{equation}
where $\hat{\sigma}, \hat{\mu}$ are mean and variance calculated over all neurons.

Interestingly, MMCosine \cite{multi5} made the assumption that vanilla fusion could all be represented as concatenation and then made adjustments to vanilla fusion methods. 
However, based on the biological theory in Sec.\ref{sec:method}, we propose the following hypothesis of linear superposition,
\vspace{-0.2cm}
\begin{equation}
\vspace{-0.2cm}
U = \zeta X_{1} + (1-\zeta) X_{2}.
\label{con:summation}
\end{equation}
$X_{1}$,$X_{2}$,$U$ are representations of two different modalities and the fusion representation respectively. Here, $\zeta$ is the adjustment weight obtained by a learnable module and has the same number of channels as $X_{1}$,$X_{2}$, and $U$. After taking the expression for minimum energy into the energy equation for the signal superimposition, the result is as follows
\vspace{-0.2cm}
\begin{equation}
\vspace{-0.2cm}
\begin{aligned}
    e_{t,U}\left(\lambda \right) = & \zeta^2 e_{t,x_1}\left(\lambda=0 \right) +(1-\zeta)^2 e_{t,x_2}\left(\lambda=0 \right) \\
    & +2\zeta(1-\zeta)E(x_1,x_2).
\end{aligned}
\label{con:basic_energy}
\end{equation}

Here we do not introduce regularization coefficients ($ \lambda$=$0$) for the two modalities separately but instead add regularization to the fused representation and overall regularization. Note, due to the non-linear operation in calculating energy, the mutual energy $E(x_1,x_2)$ here cannot be simply understood as a variable related to the linear correlation coefficient.
% Given that both $e_{t,x_1}$ and $e_{t,x_2}$ can be computed directly, essentially $e_{t,U}$ and $E(x_1,x_2)$ have the same growth relationship. So mutual energy is used instead of energy for the next calculation. 
% Then the mutual energy in Eq.\ref{con:basic_energy} can be expressed as
% \vspace{-3pt}
% \begin{equation}
% \begin{split}
% & E(x_1,x_2)= \\
% &     \frac{e_{t,U}\left(\lambda \right)-\zeta^2 e_{t,x_1}\left(\lambda=0 \right)-(1-\zeta)^2 e_{t,x_2}\left(\lambda=0 \right)}{2\zeta(1-\zeta)}.
% \end{split}
% \label{con:pro_energy}
% \end{equation}

Speaking of correlation, how to estimate modal correlations without decoupling and introducing additional classifiers is essential in this derivation. 
% The idea of using variance in the batch to represent the degree of correlation in the training data is innovative. This conclusion is drawn by constraining the correlation to observe the change in variance.
We forced the correlation based on the soft HGR maximum correlation coefficient \cite{lee2022maximal,liang2021person} of the two modalities to be exactly equal to a particular value by the MAE. Then the batch's distribution of the $\zeta$ neurons in the Eq.\ref{con:update} is observed after full training. 
A more scattered distribution of $\zeta$ is recorded under larger and smaller correlation circumstances. So we consider 
\vspace{-0.2cm}
\begin{equation}
\vspace{-0.2cm}
\begin{aligned}
% E^* = (2-tanh(k \textit{var}(\zeta))e_{t,U}-\zeta^2{e_{t,x_1}}-(1-\zeta)^2{e_{t,x_2}},
% E^* = (2-tanh(k \frac{\textit{var}(\zeta)}{\textit{var}_{max}})e_{t,U}-\zeta^2{e_{t,x_1}}-(1-\zeta)^2{e_{t,x_2}},
% E^* = &(1+k)e_{t,U}-\zeta^2{e_{t,x_1}}-(1-\zeta)^2{e_{t,x_2}},\\
r=tanh(\frac{s \sum_{i}^{} var(\zeta_i) + \lambda}{\sum_{i}^{}var_{max}(\zeta_i) + \lambda}),
\end{aligned}
\label{con:update}
\end{equation}
as an indirect measurement to estimate correlations. $var(\zeta_i)$ and $var_{max}$ are the variance of the $i$-th $\zeta$ neuron within the batch and the highest variance record of this neuron respectively. $\lambda$ is the same regulation term, usually $1e-6$ in Eq.\ref{con:regu}, and $s$ is the scaling hyperparameter. In other words, when two modalities are completely uncorrelated or completely correlated, $\zeta$ is unable to learn which modal signal is more important, resulting in a random distribution ($r\rightarrow 1$).

In a general sense, multimodal tasks are two modalities making reasonable judgments with common information. 
% This means that the multimodal neurons before fusion are somewhat correlated. 
This means both mutual energy $E(x_1,x_2)$ and overall fusion energy $e_{t,U}$ need to be optimized. But simply using $e_{t,U}$ will cause instability in extreme cases,  which makes us add an extra factor $r$ to form degradation term
\vspace{-0.2cm}
\begin{equation}
\vspace{-0.2cm}
\begin{aligned}
% E^* = (2-tanh(k \textit{var}(\zeta))e_{t,U}-\zeta^2{e_{t,x_1}}-(1-\zeta)^2{e_{t,x_2}},
% E^* = (2-tanh(k \frac{\textit{var}(\zeta)}{\textit{var}_{max}})e_{t,U}-\zeta^2{e_{t,x_1}}-(1-\zeta)^2{e_{t,x_2}},
E^* = &e_{t,U}-\zeta^2{e_{t,x_1}}-(1-\zeta)^2{e_{t,x_2}}+(1-r)e_{t,U},\\
% k &= 1 - tanh(\frac{s \sum_{i}^{} var(\zeta_i)}{\sum_{i}^{}var_{max}(\zeta_i)}),
\end{aligned}
\label{con:update}
\end{equation}
The above equation can be explained as the numerator of the transformed Eq.\ref{con:basic_energy} plus a degradation term (also analyzed in Sec.\ref{sec:fusion}) that limits the extreme conditional degeneracy of the correlation. 
% It is believed that $E^*$ is a locally optimal solution, but does not necessarily satisfy the global optimal solution.

The fusion weights $U$  are updated with the formula
\vspace{-0.2cm}
\begin{equation}
\vspace{-0.2cm}
\begin{aligned}
\bar{U}=&\operatorname{sigmoid}\left(E^*\right) \odot U,\\
% \bar{X}_{K}=&\operatorname{sigmoid} \left(\frac{1}{e_{t,x_1}\left(\lambda \right)}\right) \odot X_{K}, K \in \{1,2\}.
\end{aligned}
\label{con:final}
\end{equation}
Then, the derivation of the update of the fusion feature $U$ is completed.

Interestingly, from the perspective of uncertainty estimation, it also explains the feasibility of SimAM$^2$.
Q. Zhang \cite{uncertainty} proposed and proved an assumption that the estimated uncertainty of each modality is positively correlated with its modal-specific loss logits. Meanwhile, Zhang used energy scores calculated from the sum of logits to update the logits of each modality in late fusion.
% \vspace{-0.4cm}
% \begin{equation}
% \vspace{-0.2cm}
% \begin{aligned}
% \operatorname{Energy}\left(X_K\right)&=-\mathcal{T}_K \cdot \log \sum_n^N e^{f_n^m\left(x_{K}\right) / \mathcal{T}_K},\\
% % w^m(x)=&\alpha^m u^m(x).
% \Bar{x}_K=&\alpha_K \operatorname{Energy}\left(x_{K}\right)+\beta_K,
% \end{aligned}
% \label{con:uncertainty}
% \end{equation}
% where $\alpha_K$$<$0, $\beta_K$$>$0 are modal-specific hyperparameters. $f_n^K\left(x_K\right)$ is the output logits of classifier $f_K$ corresponding to the $n$-th class label and $\mathcal{T}_K$ is a temperature parameter. 
Based on this conclusion of Zhang, using a single variable to update the fused feature is effective in late fusion. Our proposed Eq.\ref{con:final} has the same form as Zhang proposed. So from another perspective, Eq.\ref{con:final} can be seen as considering the uncertainty of each neuron more meticulously, rather than the whole feature as \cite{uncertainty}. 
% Therefore, we also applied regulation of the fusion weights from the historical training trajectory \cite{uncertainty} to limit growth.

% Below is an example of how to insert images. Delete the ``\vspace'' line,
% uncomment the preceding line ``\centerline...'' and replace ``imageX.ps''
% with a suitable PostScript file name.
% -------------------------------------------------------------------------
\vspace{-0.3cm}

\subsection{Improvement of Vanilla Fusion}
\label{sec:fusion}
\vspace{-0.1cm}
Total 3 methods of vanilla fusion are discussed and 2 of them improved on the grounded theory of SimAM$^2$ (Sec.\ref{sec:att}).

\textbf{Summation}: There are two dominant methods of summation widely used. The simplest method is the direct summation like \cite{multi3}, which does not adjust $\zeta$ automatically. So the implementation of SimAM$^2$ is just letting $\zeta$=$\textbf{0.5}$ in Eq.\ref{con:summation}. As a second and slightly more complex approach, learnable parameters are used to adjust the weights of both sides (learnable $\zeta$ in Eq.\ref{con:summation}) \cite{av_asso1}. 
% $\zeta$ is the adjustment weight obtained through a fully connected layer on top of the concatenation of each unimodal feature. 
In practice, the latter method seems to be more advantageous apart from being more computationally intensive.

\textbf{FiLM}: An adaptive affine fusion proposed by FiLM \cite{film} innovatively introduces learnable feature modulation, but obviously using different major modality will have a large impact on the results.
\vspace{-0.2cm}
\begin{equation}
\vspace{-0.2cm}
\begin{aligned}
\operatorname{FiLM}\left(F_{i, c} \mid \gamma_{i, c}, \beta_{i, c}\right)=\gamma_{i, c} F_{i, c}+\beta_{i, c}. 
\label{con:film}
\end{aligned}
\end{equation}
%\vspace{-0.2cm}
FiLM learns functions $f$ and $h$ which have input of $x_i$ and output $\gamma_{i, c}=f_c\left(\boldsymbol{x}_i\right)$ and $\beta_{i, c}=h_c\left(\boldsymbol{x}_i\right)$. In other words, swapping the order of the two modalities in the calculation will give different results. So $\gamma_{i, c}$ and $\beta_{i, c}$ modulate a neural network’s activations $F_{i, c}$, whose subscripts refer to the $i^{th}$ input’s $c^{th}$ feature. Thinking about it another way in fact the first term of Eq.\ref{con:film} is just a coupling of one modality on top of another modality weight. So the two terms of Eq.\ref{con:film} still essentially have completely different modality information. At the same time, for more general multimodal domains, we still want to use $\zeta$ to balance the training process. So Eq.\ref{con:film} is transformed into the form of Eq.\ref{con:summation} as
\vspace{-0.2cm}
\begin{equation}
\vspace{-0.2cm}
\begin{aligned}
\operatorname{FiLM}\left(F_{i, c} \mid \gamma_{i, c}, \beta_{i, c}\right)=\zeta \gamma_{i, c} F_{i, c}+(1-\zeta)\beta_{i, c}.
\label{con:film_plus}
\end{aligned}
\end{equation}
After treating these terms as different modalities, the rest of the calculation is the same as the summation. Later experiments demonstrated that the introduction of $\zeta$ alleviated the problem generated by different major modalities.

\textbf{Concatenation}: It is obvious that concatenation can be seen as the case where two vectors are respectively padded at different positions and expressed as Eq.\ref{con:summation}. However, the vectors after padding are clearly linearly uncorrelated. 
% So the degradation term in Eq.\ref{con:pro_energy} will degrade to $0$ as analysis in Sec.\ref{sec:result} and Fig.\ref{fig:gated}. 
So the degradation term and the mutual energy in Eq.\ref{con:basic_energy} will degrade to $0$ as analysed in Sec.\ref{sec:result}. The reason why the mutual energy is not completely equal to zero is that there have been nonlinear transformations on the features before.
% So, under this condition, the SimAM$^2$ method eventually degraded to just the attention of SimAM for the two modalities separately.

\vspace{-0.3cm}

\subsection{Decoupling-free gradient modulation}
\label{sec:Non-decoupled}
\vspace{-0.1cm}
A number of works \cite{multi3,multi5} have emerged in recent years to provide solutions to the imbalance of multimodal training from the perspective of metric learning, and backpropagation respectively. Let's take OGM-GE \cite{multi3} as an example. OGM-GE needs to decouple the fused features and then calculate the logits of each modality. However, the fusion involves non-linear layers such as sigmoid, and more complex combinations of multi-layer convolution and activation layers, which makes it hard to accurately calculate the logits of individual modality. So, the final result is not as good as it could be.
% The terms used in the OGM-GE
% \vspace{-0.2cm}
% \begin{equation}
% \vspace{-0.2cm}
% \begin{aligned}
% & s_i^U=\sum_{k=1}^M 1_{k=y_i} \cdot \operatorname{softmax}\left(W_t^U \cdot \varphi_t^U\left(\theta^U, x_i^U\right)+\frac{b}{2}\right)_k, \\
% % & s_i^v=\sum_{k=1}^M 1_{k=y_i} \cdot \operatorname{softmax}\left(W_t^v \cdot \varphi_t^v\left(\theta^v, x_i^v\right)+\frac{b}{2}\right)_k,
% \end{aligned}
% \end{equation}
% to calculate the discrepancy ratios, $\rho_t^v$ and $\rho_t^a$, are essentially decoupling and scoring  of the fused features. 

According to OGM-GE, the discrepancy ratios are used to monitor the discrepancy of their contribution to the learning objective, which exactly fits our definition of the adjustment weight $\zeta$. So, naturally, we propose the calculation of the discrepancy ratio without decoupling the fused features as
\vspace{-0.2cm}
\begin{equation}
\vspace{-0.2cm}
\rho_t^v=\frac{\sum_{i} \zeta_i}{\sum_{i}1-\zeta_i}.
\end{equation}
$\rho_t^a$ is accordingly defined as the reciprocal of $\rho_t^v$. The rest of the propagation remains the same as for the OGM-GE.
To analyze the efficiency of the two methods, we performed a correlation analysis of the discrepancy ratios obtained from the two different calculation methods during the ablation experiments. It was found that the values from both methods are indeed strongly correlated. 
So, avoiding the accumulation of decoupling errors brings optimization to gradient modulation.
% Moreover, our method has shown more competitiveness in the datasets by avoiding the accumulation of decoupling errors.

\vspace{-0.4cm}

\section{Experiment}
\label{sec:expe}

\vspace{-0.2cm}

Two mainstream audio-visual classification datasets were used to ensure the method's effectiveness. CREMA-D \cite{dataset1} is an audio-visual dataset utilizing face and voice data to classify six basic emotional states, which contain 7,442 video clips of 2-3 seconds from 91 actors speaking several short words. This dataset consists of the 6 most usual emotions. VGGSound \cite{dataset2} is a large-scale video dataset that contains 309 classes, covering a wide range of audio events in everyday life. The duration of each video is 10 seconds, and the division of the dataset is the same as proposed in \cite{dataset2}. For CREMA-D and VGGSound, we employ the exact same data-preprocessing, overall backbones, and losses as OGM-GE \cite{multi3} for Table.\ref{table:OGM-GE}. As for additional experiments, we insert our module into 3 representative frameworks AGVA \cite{AVE1} and PSP \cite{AVE2} on AVE dataset \cite{AVE1}, and FOP\cite{av_asso1} on Seen-Heard/Unseen-Unheard dataset \cite{dataset3} with adjustment of very few parameters to explore the impacts.

For each task, the parameter $s$ in Eq.\ref{con:update} is roughly set to around $2.5$ without elaborate searching. And all the experiments are conducted on 4 NVIDIA 3090Ti GPUs.
\vspace{-0.4cm}

\section{Result and Analysis}
\label{sec:result}

\vspace{-0.2cm}

\begin{table}[]
    \centering
    \captionsetup{font={small}}
    \caption{
    \label{table:OGM-GE}
    \small Performance of multimodal classification}
    \vspace{-0.2cm}
    \scalebox{0.6}{
\begin{threeparttable}
\begin{tabular}{cccccc}
\hline
\multirow{2}{*}{Method}                                                        & \multirow{2}{*}{Fusion} & \multicolumn{2}{c}{CREMA-D} & \multicolumn{2}{c}{VGGSound} \\
                                                                         &                          & Top1-Acc(\%)      & mAP(\%)      & Top1-Acc(\%)       & mAP(\%)       \\
\hline
\multirow{3}{*}{Baseline}                                                & Concatenation$\dagger$            & 51.7              & 53.5    & 49.1               & 52.5     \\
                                                                         & Summation                & 51.5              & 53.5    & 49.1               & 52.4     \\
                                                                         & FiLM                     & 50.6              & 52.1    & 48.5               & 51.6     \\ 
\hline
\multirow{3}{*}{+SimAM$^2$}                                                   & Concatenation$\dagger$            & 51.3                 & 53.0       & 49.2                  & 52.7        \\
                                                                         & Summation                & 54.3                 & 55.7       & 49.3                  & 52.9        \\
                                                                         & FiLM                     & 52.4                 & 54.1       & 48.9                  & 51.2        \\
\hline
\multirow{3}{*}{+OGM-GE\cite{multi3}}                                                 & Concatenation$\dagger$            & 61.9              & 63.9    & 50.6               & 53.9     \\
                                                                         & Summation                & 62.2              & 64.3    & 50.4               & 53.6     \\
                                                                         & FiLM                     & 55.6              & 57.4    & 50.0               & 52.9     \\
\hline
\multirow{3}{*}{\begin{tabular}[c]{@{}c@{}}+OGM-GE\\ +SimAM$^2$\end{tabular}} & Concatenation$\dagger$            & 62.1                & 64.0       & 51.0                  & 54.2        \\
                                                                         & Summation                & 63.7                 & 65.7       & 51.4                 & 54.0        \\
                                                                         & FiLM                     & 58.6                 & 60.3       & 50.1                  & 53.1       \\
\hline
\multirow{3}{*}{\begin{tabular}[c]{@{}c@{}}+Decouping-Free\\ +SimAM$^2$*\end{tabular}} & Concatenation$\dagger$            & 60.1                 & 61.7       & 50.1                  & 53.3        \\
                                                                         & Summation                & \textbf{63.9}                 & \textbf{66.1}       & \textbf{51.6}                 & \textbf{54.4}        \\
                                                                         & FiLM                     & \textbf{59.8}                 & \textbf{61.4}       & \textbf{50.4}                  & \textbf{53.}5       \\
% \hline
% \multirow{3}{*}{+OGM-GE*}                                                 & Concatenation            & 67.6              & 69.3    & 60.8               & -     \\
%                                                                          & Summation                & 66.9              & 68.1    & 60.4               & -     \\
%                                                                          & FiLM                     & 67.8              & 69.5    & 50.0               & -     \\
% \hline
% \multirow{3}{*}{\begin{tabular}[c]{@{}c@{}}+OGM-GE*\\ +Ours\end{tabular}}                                                 & Concatenation            & 69.9              &  70.9   & 61.0               & -     \\
%                                                                          & Summation                & 70.2              &   71.3   & 61.3               & -     \\
%                                                                          & FiLM                     & 70.8              &   72.0  & 60.9               & -     \\
 \hline
\end{tabular}
\begin{tablenotes}
\normalsize
\item[*] To make clear comparison, learnable $\zeta$ is only applied to ‘‘+Decouping-Free + SimAM$^2$" experiment.
\item[$\dagger$] The experiments on concatenation are only used to show the effect on the results in the most extreme scenario.
\end{tablenotes}
\end{threeparttable}
}
\vspace{-0.6cm}
\end{table}

\textbf{Comparison of Classification}: SimAM$^2$ based on Sec.\ref{sec:fusion} was applied directly to the CREMA-D and VGGSound, and the improvement of 2\% on average in accuracy was obtained under the conditions of two different vanilla fusions. We first conduct experiments on the baseline where the two modalities are directly co-trained. The same problem of the unbalanced problem as described in \cite{multi3} also occurs in the baseline training process. So, likewise, OGM-GE is introduced to equilibrium convergence to explore more comparisons. Finally, it is observed in the experiment that the best results are obtained if the decoupling-free strategy is applied along with SimAM$^2$. The comparison of the three vanilla fusion experiments on the CREMA-D and VGGSound shows that SimAM$^2$ can indeed introduce more robustness in the fusion process. Clearly, since the design of SimAM$^2$ is based on signal superposition theory, it can be observed that SimAM$^2$ has a better performance gain in terms of direct summation. However, SimAM$^2$ is less of a boost for concatenation, due to the fact that the concatenation method is essentially a simple summation of two strictly orthogonal signals as described in Sec.\ref{sec:fusion}.

% \begin{figure}[t]
% \centering
% \includegraphics[width=8cm]{figs/dr.png}
% % \vspace{-0.2cm}
% \caption{
% \small
% The discrepancy ratio was generated by 2 different methods.}
% \label{fig:dr}
% \vspace{-0.4cm}  %调整图片与上文的垂直距离
% \end{figure}

\textbf{Comparison of decoupling-free gradient modulation}: As mentioned above, for the orthogonal reasons described in Sec.\ref{sec:fusion}, the discrepancy ratio obtained using decoupling-free is not accurate when using the concatenation method for fusion. However, it is worth mentioning that the concatenation method is the easiest method in OGM-GE to decouple. So we prefer to keep the decoupling method of OGM-GE in the concatenation method, but the remaining two fusion methods still demonstrate stronger robustness in the case of decoupling-free methods. In order to further analyze the relationship between the discrepancy ratios obtained by the two methods, we calculate and record the two ratios at the same time
% we plot the two ratios after smoothing into  Fig.\ref{fig:dr}
and observe a clear correlation. Again, it can be concluded from the experimental results that the non-decoupled method is an optimization of the decoupled method.

% \usepackage{multirow}
\begin{table}[]
    \centering
    \captionsetup{font={small}}
    \caption{
    \label{table:AVE}
    \small Performance of Audio-visual Event Localization}
    \vspace{-0.2cm}
    \scalebox{0.67}{
\begin{tabular}{lll}
\hline
Method                                                                                                & Network & Acc(\%)   \\ \hline
\multirow{2}{*}{baseline}                                                                             & AVGA\cite{AVE1}    & 72.0 \\
                                                                                                      & PSP\cite{AVE2}     & 76.2 \\ \hline
\multirow{2}{*}{+SimAM$^2$}                                                            & AVGA    & 72.4 \\
                                                                                                      & PSP     & 76.7 \\ \hline
\multirow{2}{*}{+OGM-GE}                                                                              & AVGA    & 72.8 \\
                                                                                                      & PSP     & 76.9 \\ \hline
\multirow{2}{*}{\begin{tabular}[c]{@{}l@{}}+SimAM$^2$\\ +Decoupling-Free\end{tabular}} & AVGA    &   \textbf{73.0}   \\
                                                                                                      & PSP     & \textbf{77.2}\\ \hline    
\end{tabular}
}
\vspace{-0.3cm}
\end{table}

\textbf{Comparison of fusion in other applications}: 
In order to further validate the versatility of SimAM$^2$ in more general scenarios, the fusion of SimAM$^2$ is applied in multiple tasks based on the idea of plug-and-play. Firstly, applying the SimAM$^2$ method to audio-visual event localization \cite{AVE1,AVE2} in late fusion, it is observed that general performance gains are achieved regardless of the network used in Table.\ref{table:AVE}. Another interesting experiment is on FOP \cite{av_asso1} in the face-voice association. This simple modification of fusion also seems to yield slightly better results on unseen-unheard and seen-heard datasets \cite{dataset3} in Table.\ref{table:FOP} as well. So, the improvement brought by the SimAM$^2$ fusion method demonstrates that our method still generally brings consistent performance for fusion.
% \usepackage{multirow}

\begin{table}[]
    \centering
    \captionsetup{font={small}}
    \caption{
    \label{table:FOP}
    \small Performance of Face-Voice Association}
    \vspace{-0.2cm}
\scalebox{0.67}{
\begin{threeparttable}
\begin{tabular}{lllll}
\hline
\multirow{2}{*}{Methods} & \multicolumn{2}{l}{Seen-Heard}                 & \multicolumn{2}{l}{Unseen-Unheard}                              \\
                         & EER(\%)           & AUC(\%)                             & EER(\%)                             & AUC(\%)                             \\
\hline
% DIMNet                   & -             & -                              & 24.9                           & -                              \\
% Learnable Pins           & 21.4          & 87.0                           & 29.6                           & 78.5                           \\
% MAV-Celeb                & -             & -                              & 29.0                           & 78.9                           \\
% Single Stream Network    & 17.2          & 91.1                           & 29.5                           & 78.8                           \\
FOP\cite{av_asso1}                      & 19.3          & 89.3                           & 24.9                           & 83.5                           \\
% \hline
FOP+SimAM$^2$*                & \textbf{18.5} & \textbf{89.8} & \textbf{24.6} & \textbf{83.7}\\
\hline
\end{tabular}
\begin{tablenotes}
    \normalsize
\item[*] Learnable $\zeta$ has already been applied to original FOP.
\end{tablenotes}
\end{threeparttable}
}
\vspace{-0.6cm}
\end{table}

\vspace{-0.4cm}
\section{conculsion}
\label{sec:con}

\vspace{-0.2cm}

In this paper, a novel attention method based on informatics energy theory is reasoned. Then based on this approach, the vanilla fusion is improved along with the decoupling-free gradient modulation scheme, allowing a general improvement of multimodal tasks on different datasets and methods. 
% Moreover, a possible understanding based on uncertainty have been interpreted to prove the theoretical feasibility of the method. 
In the future, provable fusion based on informatics theory needs to be discovered.

% \begin{figure}[htb]

% \includegraphics[width=\linewidth]{figure.pdf}
% %
% \caption{\textbf{Figure and table captions} should be 10-point Helvetica (or a similar sans-serif font, e.g. 9-point Arial), boldface. Callouts should be 9-point Helvetica, non-boldface. Initially capitalize only the first word of each figure caption and table title. Figures and tables must be numbered separately. For example: “Figure 1. Database contexts”, “Table 1. Input data”. Figure captions are to be below the figures. Table titles are to be centered above the tables. Use the “Caption” style for figure captions. The right style is automatically chosen when you right-click on the picture and select from the pop-up menu “Caption”. This ensures automatic figure numbering, which also can be cross-referenced in the text.}
% \label{fig:res}
% %
% \end{figure}

%\vfill
%\pagebreak

\vfill\pagebreak

\begin{spacing}{0.9}
\bibliographystyle{IEEEbib}
\small
\bibliography{refs}
\end{spacing}

\end{document}